%% file: folded_it.tex
\documentclass[onecolumn,11pt]{IEEEtran}
\usepackage[cmex10]{amsmath}
\usepackage{amssymb, amsfonts, mathrsfs}

\usepackage{hyperref,url}
\usepackage{theorem}

\theoremstyle{plain}
\newtheorem{theorem}{Theorem}{\bfseries}{\it}
\newtheorem{definition}{Definition}{\bfseries}{\it}
\newtheorem{proposition}{Proposition}{\bfseries}{\itshape}
\newtheorem{lemma}{Lemma}{\bfseries}{\itshape}
\newtheorem{corollary}{Corollary}{\bfseries}{\itshape}
\newtheorem{notation}{Notation}{\bfseries}{\itshape}

\theoremstyle{remark}
\newtheorem{remark}{Remark}{\bfseries}

\newcommand\QM{\ensuremath{{\rm QM}}}
\newcommand\QD{\ensuremath{{\rm QD}}}
\newcommand\QC{\ensuremath{{\rm QC}}}

\newcommand{\eqdef}{\stackrel{\text{def}}{=}}

\newcommand{\F}{\ensuremath{\mathbb{F}}}
\newcommand{\G}{\ensuremath{\mathbb{G}}}

\newcommand{\Z}{\ensuremath{\mathbb{Z}}}

\newcommand{\fq}{\F_q} 
\newcommand{\Fq}{\ensuremath{\mathbb{F}_q}}

\newcommand{\Fqm}{\ensuremath{\mathbb{F}_{q^m}}}
 
\newcommand{\fqm}{\F_{q^m}}

\DeclareMathOperator{\tr}{Tr}

\newcommand{\word}[1]{\ensuremath{\boldsymbol{#1}}}  
\newcommand{\mat}[1]{\ensuremath{\boldsymbol{#1}}}
\newcommand{\code}[1]{\ensuremath{\mathscr{#1}}}

\newcommand{\goppa}{\ensuremath{\code{G}}}
\newcommand{\pg}{\Gamma}

\newcommand{\Gm}{\mat{G}}

\newcommand{\cv}{\word{c}}
\newcommand{\gv}{\word{g}}
\newcommand{\xv}{\word{x}}
\newcommand{\yv}{\word{y}}
\newcommand{\zv}{\word{z}} 

\newcommand{\uv}{\word{u}}

\renewcommand{\leq}{\leqslant} 
\renewcommand{\le}{\leqslant} 
\renewcommand{\geq}{\geqslant} 
\renewcommand{\ge}{\geqslant} 

 \newcommand{\JPAlt}{\mathcal{A}}

 \newcommand{\GRS}[3]{\text{\bf GRS}_{#1}(#2,#3)}
 \newcommand{\Alt}[3]{\code{A}_{#1}(#2,#3)}
 \newcommand{\Goppa}[2]{\code{G}\left(#1,#2\right)}
 \newcommand{\mAlt}[3]{\JPAlt_{#1}(#2,#3)}
 \newcommand{\mgoppa}[2]{\goppa(#1,#2)}

 \newcommand{\cond}[2]{\overline{#2^{#1}}}
 \newcommand{\csigma}[2]{\widetilde{#2}^{#1}}
 \newcommand{\Cc}{\code{C}}
\newcommand{\Oscr}{\code{O}}

\begin{document}   
   

\title{Folding Alternant and Goppa Codes with Non-Trivial Automorphism Groups
\footnote{A preliminary version of this paper will be presented at ISIT'14 under the title  "Structural Weakness of Compact Variants of the McEliece Cryptosystem".}
}

\author{%
\IEEEauthorblockN{Jean-Charles Faug\`ere \IEEEauthorrefmark{2}\IEEEauthorrefmark{1}\IEEEauthorrefmark{3},
Ayoub Otmani\IEEEauthorrefmark{4},
Ludovic Perret\IEEEauthorrefmark{1}\IEEEauthorrefmark{2}\IEEEauthorrefmark{3},
Fr\'ed\'eric de Portzamparc \IEEEauthorrefmark{1}\IEEEauthorrefmark{2}\IEEEauthorrefmark{3}\IEEEauthorrefmark{6} and
Jean-Pierre Tillich\IEEEauthorrefmark{5}}

\IEEEauthorblockA{\IEEEauthorrefmark{1}Sorbonne Universit\'{e}s, UPMC Univ Paris 06, POLSYS, UMR 7606, LIP6, F-75005, Paris, France\\ludovic.perret@lip6.fr} 

\IEEEauthorblockA{\IEEEauthorrefmark{2} INRIA, Paris-Rocquencourt Center, \\
jean-charles.faugere@inria.fr}

\IEEEauthorblockA{\IEEEauthorrefmark{3} CNRS, UMR 7606, LIP6, F-75005, Paris, France}

\IEEEauthorblockA{\IEEEauthorrefmark{4}
Normandie Univ, France; UR, LITIS, F-76821 Mont-Saint-Aignan, France.\\ayoub.otmani@univ-rouen.fr.}

\IEEEauthorblockA{\IEEEauthorrefmark{5}INRIA, Paris-Rocquencourt Center, \\ jean-pierre.tillich@inria.fr}

\IEEEauthorblockA{\IEEEauthorrefmark{6}Gemalto, 6 rue de la Verrerie
  92190, Meudon, France \\ frederic.urvoydeportzamparc@gemalto.com}
}
\maketitle

 \begin{abstract}
The main practical limitation of the McEliece public-key encryption scheme is probably the size of its key. A famous trend to overcome this issue  is to focus  on subclasses of alternant/Goppa codes
with a non trivial automorphism group. Such codes display then \textit{symmetries} allowing
 compact parity-check or generator matrices. For instance, a key-reduction is obtained by taking 
 {\it quasi-cyclic} (\QC{}) or {\it quasi-dyadic} (\QD{}) alternant/Goppa codes.
 We show that the use of such  \textit{symmetric} alternant/Goppa  codes in cryptography introduces a fundamental weakness. It is indeed possible to reduce the key-recovery  on the original symmetric public-code to the key-recovery on a (much) smaller code that has not anymore symmetries. This result is obtained thanks to a new operation on codes  called \textit{folding}  that exploits the knowledge of the automorphism group. This operation consists in adding the coordinates of codewords which belong to the same orbit under the action of the automorphism group. The advantage is twofold: the  reduction factor can be as large as the size of the orbits, and it preserves a fundamental property: 
folding the dual of an alternant (\textit{resp}. Goppa) code provides the dual of an alternant  (\textit{resp}. Goppa) code. A key point is to show that all the existing constructions of alternant/Goppa codes with symmetries follow a common principal of taking codes whose support is globally invariant 
 under the action of affine transformations (by building upon prior works of T. Berger and  A. D{\"{u}}r). This enables not only to present a unified view but also to generalize the construction of \QC{}, \QD{} and even \textit{quasi-monoidic} (\QM{}) Goppa codes. All in all, our results can be harnessed to boost up any key-recovery attack on McEliece systems based on symmetric alternant or Goppa codes, and in particular algebraic attacks.
 \end{abstract}

\input{introduction.tex}

\input{preliminaries.tex}

\input{symmetries.tex}

\input{invariant_polynomial}
\input{reduction}

\input{implications}

\bibliographystyle{IEEEtran}
\bibliography{IEEEabrv,crypto}

\newpage 

\appendix

\input{proofs}

\end{document}

%% file: introduction.tex
\section{Introduction}

Some significant research efforts have been put recently in code-based cryptography to reduce by  a large factor the public key sizes.
This has resulted in keys which are now only  a few times larger than RSA keys (see \cite{BLM11a,MTSB13} for instance).
 This is obtained by  focusing  on 
codes having \emph{symmetries}, that is to say, codes  having 
a non-trivial automorphism group. 
Such codes have the advantage of admitting a compact parity-check or generator matrix \cite{Gaborit05,BCGO09,MB09,BLM11a,Per12}. Quasi-cyclic (\QC{}) codes represent a good example of the use of symmetries in cryptography 
to build public-key encryption schemes with short keys \cite{Gaborit05,BCGO09}. It was then followed by a series of papers proposing alternant and Goppa codes with different automorphism groups like quasi-dyadic (\QD{}) Goppa or Srivastava codes \cite{MB09,Per12} and 
quasi-monodic (\QM{}) codes \cite{BLM11a}.
The rationale behind this is the fact that the additional structure does not deteriorate the security of the cryptographic scheme.  
This hope was eroded by the apparition of specific attacks \cite{OTD08,OTD10}
and algebraic attacks \cite{FOPT10,FOPT_SCC10,UL09} against \QC{}/\QD{} alternant/Goppa codes.  
Despite these preliminary warning signals, the design of compact McEliece schemes remains a rather popular topic of 
research e.g. \cite{qd2012,BLM11a,Per12,BCMN10,Barb11}. 
Besides these cryptographic motivations, the search for Goppa codes, and more generally alternant codes, with 
non-trivial automorphisms is in itself an important issue in coding theory. Several papers focused on the problem of 
constructing quasi-cyclic Goppa codes \cite{bb00,ryanfitz04}, or identifying alternant and Goppa codes invariant under a given 
permutation \cite{Berg99,Ber00b,Ber00a}.

\subsection*{Main Results}
All the constructions of \emph{symmetric} alternant/Goppa codes presented in previous works might look at first glance 
unrelated, like \emph{ad hoc} constructions designed for a very specific goal. In \cite{MB09} symmetric \QD{} Goppa codes are constructed by 
using the narrower class of separable Goppa codes which have all their roots of multiplicity one in the field over which the coefficients of the Goppa polynomial
are taken and by choosing these roots in an appropriate manner; the same approach is followed to obtain more general \QM{} Goppa codes in \cite{BLM11a}, whereas in
\cite{BCGO09} the authors rely on the larger class of alternant codes to obtain a large enough family of \QC{} codes in a  McEliece like scheme.
Building upon the work of \cite{Dur87,Ber00a,Ber00b}, we show in this paper that all the \QC{}, \QD{} and \QM{} alternant/Goppa codes which are constructed in \cite{BCGO09,MB09,BLM11a} 
rely actually on a common principle (Proposition~\ref{pr:description}). They are 
all equipped with non-trivial  automorphism groups that involve some affine transformations  leaving globally invariant their support.
This property imposes on the non-zero scalars defining the alternant codes the constraint of being built from a root of unity. In the case of Goppa codes, 
this constraint is translated into a \emph{functional equation} of the form $\alpha \pg(az+b) = \pg(z)$ that the Goppa polynomial $\pg(z)$ has to satisfy, where $\alpha$ is a root of unity and $a$, $b$ belong to the underlying finite field on which the support is defined. 
We fully characterize polynomials satisfying such equation in Proposition~\ref{pr:PolynomialForm}. This enables not only to present a unified view but also to generalize the construction of \QC{}, \QD{} and \QM{} Goppa codes (Proposition~\ref{pr:monoidic_alternant}). In particular, there is no need to use separable polynomials like in \cite{MB09} for getting \QD{} Goppa codes.
Notice that this will also show that it is in principle not compulsory to take the larger family of alternant codes instead of Goppa codes as in \cite{BCGO09} to obtain a large enough family of 
\QC{} codes in a McEliece scheme: in fact there is nothing special with respect to \QD{} Goppa codes instead of \QC{} Goppa codes because there are roughly as many
\QD{} Goppa codes as there are \QC{} Goppa codes (for a same size of automorphism group) with our way of constructing them.


The major contribution of our paper is to prove that alternant and Goppa codes with symmetries can be seen as an \emph{inflated} version of a smaller alternant code \emph{without} symmetries. We  call this latter a \emph{folded} code because we show that it can be obtained easily by adding the coordinates which belong to the same 
orbit under the action of a permutation of the automorphism group. More importantly, we can also express precisely the relationship between the supports and the non-zero 
scalars defining the alternant/Goppa with symmetries and their associated folded codes. These links are so explicit for the non-zero scalars that knowing those of the folded code is sufficient for knowing those of the original symmetric alternant/Goppa codes.
These results have an important impact in cryptography. First the length and the dimension of the folded code is generally divided by the cardinality of the automorphism group. 
It means in particular that the use of compact alternant/Goppa 
codes introduces a fundamental weakness: decreasing the size of the public-key 
as in \cite{BCGO09,MB09,BLM11a} necessarily implies a deterioration of the security. Furthermore, since the non-zero scalars of the folded code bear crucial
 information, it then allows in the context of algebraic attacks as proposed in (\cite{FOPT10,FOPT_SCC10,FOPPT13}), to reduce a key-recovery  attack on the original 
 public-code to the one on a smaller code, that is to say with less variables in the polynomial system. For instance, we can reduce the key-recovery of a 
 quasi-dyadic Goppa code of length $8192$ and dimension $4096$ to the key-recovery on a Goppa code of length $64$ and dimension $32$.  

Interestingly enough, the folded code, if used in a McEliece-like encryption scheme, would have the same key size as the original scheme but without symmetries.
In other words, the very reason which allowed to reduce the key size in \cite{BCGO09,MB09,BLM11a,BCMN10} can be used to derive a \emph{reduced} McEliece scheme whose key-recovery hardness and key size is equivalent to the original system. 

\subsection*{Comparison with ``Structural Cryptanalysis of McEliece Schemes with Compact Keys" \cite{FOPPT13}}
This paper is a companion paper of \cite{FOPPT13} which has been submitted separately.
In \cite{FOPPT13}, we mainly focused on the cryptanalysis of \QM{} Goppa codes. That is, we \cite{FOPPT13} developed new algebraic tools for solving the algebraic systems arising in the cryptanalysis \QM{} Goppa codes, reported various experimental results and prove in addition partial 
results on folded \QM{} Goppa codes. In this submission, we present a much deeper
and more systematic treatment of the the folding process. 
In \cite{FOPPT13}, the folding was performed directly over \QM{} Goppa codes and it was proved there that it results in 
a subcode of a Goppa code of reduced length. Using a slightly different approach (by considering the dual of the codes),
 we obtain here a much stronger result which holds
in a more general setting. Namely, we 
 prove  that if we perform folding on the dual of  \QC, \QD{} or \QM{} affine induced Goppa/alternant codes  (this applies for instance to all the codes constructed in 
\cite{BCGO09,MB09,BCMN10,BLM11a}) we obtain a reduced dual Goppa or alternant code where the reduction factor can be as large as the size of the cyclic or monodic blocks of a symmetric
parity-check matrix attached to these codes. 
Folding preserves here the structure of the dual code: if we start with the dual of an alternant code we end up with the dual of an alternant code and if we start with the dual of  a Goppa code
we end up with the dual of a Goppa code.

%% file: preliminaries.tex
\section{Alternant and Goppa Codes}

In this section we introduce notation which is used in the whole paper and recall
a few well known facts about alternant and Goppa codes.
Throughout the paper, the finite field of $q$ elements with $q$ being
a power of a prime number $p$ is denoted by $\Fq$. Vectors are denoted by bold letters like $\xv$ and the notation $\xv=(x_i)_{0 \leq i <n}$ or $\xv=(x_i)_{i=0}^{n-1}$ will be used in some cases.
The ring of polynomials with coefficients in a finite field $\F$  
is denoted by $\F[z]$, 
while the subspace of $\F[z]$ of polynomials of degree less than $t$ 
(\textit{resp.} less than or equal to $t$)
is denoted by $\F[z]_{<t}$ (\textit{resp.} $\F[z]_{\leq t}$). 
When $\xv=(x_i)_{0 \leq i <n}$ is a vector in $\F^n$ and  $Q(z)$ is a polynomial in $\F[z]$,  $Q(\xv)$ stands for $\left(Q(x_0),\dots, Q(x_{n-1}) \right)$. 
In particular for any vector $\uv = (u_0, \dots,u_{n-1})$ and for all $a , b\in\F$ then $a\uv+b$ stands for the vector $(au_0+b, \dots ,
au_{n-1}+b)$.


\begin{definition}[Generalized Reed-Solomon code] \label{def:defGRS}
Let $q$ be a prime power and
$k$, $n$ be integers such that $1 \leqslant k < n \leqslant q$.
Let $\xv$ and $\yv$ be two $n$-tuples such that the entries of $\xv$ are
pairwise distinct elements of $\fq$  and those of $\yv$ are nonzero elements in $\fq$.
The generalized Reed-Solomon code $\GRS{k}{\xv}{\yv}$ of dimension $k$ 
is the $k$-dimensional vector space: 
\[
\GRS{k}{\xv}{\yv} \eqdef 
\Big\{
\big(y_0P(x_0),\dots{},y_{n-1}P(x_{n-1}) \big)  \mid P \in \fq[z]_{< k}
\Big \}.
\]
\end{definition}

A useful property of these codes is given in \cite[Chap. 12, \S 2]{MacSloBook}.

\begin{proposition}\label{prop:dualGRS} 
Keeping the notation of Definition~\ref{def:defGRS},  there exists a vector $\zv \in \fq^n$ 
such that $\GRS{k}{\xv}{\yv}^{\perp} = \GRS{n-k}{\xv}{\zv}$.
\end{proposition}

This leads to the definition of alternant codes.

\begin{definition}[Alternant code, degree, support, multiplier] \label{def:subfield_subcode}
Let $\xv, \yv \in \fqm^n$ be two vectors such that the entries of $\xv$
are pairwise distinct and those of $\yv$ are all nonzero, and let $r$ and $m$ be positive integers.
The alternant code $\Alt{r}{\xv}{\yv}$ defined over $\fq$ 
is the \emph{subfield subcode} over $\Fq$ of 
$\GRS{r}{\xv}{\yv}^{\perp} \subset \fqm^n$: 
$$
\Alt{r}{\xv}{\yv} \eqdef \GRS{r}{\xv}{\yv}^{\perp} \cap \fq^n.
$$
The integer $r$ is the {\em degree} of the alternant code,
$\xv$ is a {\em support} and $\yv$ is 
a {\em multiplier} of the alternant code.
\end{definition}


 The
dual of a subfield subcode is known to be a trace code \cite{Del75a}. 
From this it follows that

\begin{lemma}\label{lem:trace}
The dual $\mAlt{r}{\xv}{\yv}^\perp$ of the alternant code $\mAlt{r}{\xv}{\yv}$ of degree $r$ and extension $m$ 
over $\fq$ is given by:
\begin{equation*} \label{eq:trace}
\mAlt{r}{\xv}{\yv}^\perp = \tr \Big( \GRS{r}{\xv}{\yv} \Big) =
\Big\{ 
\big( \tr(c_0),\dots{},\tr(c_{n-1}) \big) \mid (c_0,\dots{},c_{n-1}) \in \GRS{r}{\xv}{\yv} 
\Big \}
\end{equation*}
where $\tr$ is the trace map from $\fqm$ to $\fq$ defined by
$\tr(z) = z + z^q + \dots + z^{q^{m-1}}$.
\end{lemma}

Let us remark that an alternant code has many equivalent descriptions as shown by the following proposition whose proof can be found in \cite[Chap. 10, p. 305]{MacSloBook}.

\begin{proposition} \label{prop:affinepreserve}
For all $a \in \F_{q^m} \setminus \{0\}$, $b \in \F_{q^m}$, and $c \in \F_{q^m} \setminus \{0\}$, 
it holds that:
\[
\Alt{r}{\xv}{\yv} =\Alt{r}{a \xv + b}{c \yv}.
\]
\end{proposition}

We introduce now Goppa codes which form an important  subfamily of alternant codes.

\begin{definition}[Classical Goppa codes]

Let $\xv=(x_0,\dots,x_{n-1})$ be an $n$-tuple of distinct elements
of $\fqm$ and 
choose $\pg(z) \in \Fqm[z]$ of degree $r$  
such that $\pg(x_i) \neq 0$ for all $i \in \{0,\dots{},n-1 \}$.
The Goppa code $\code{G}(\xv,\pg)$ of degree $r$ over $\F_q$ associated to $\pg(z)$
is the alternant code $\code{A}_r(\xv,\yv)$ with
\[
y_i = \frac{1}{\pg(x_i)}.
\]
$\pg(z)$ is called the {\em Goppa polynomial} and
$\xv$ is the {\em support} 
of the Goppa code.
\end{definition}

%% file: symmetries.tex
\section{Construction of  Symmetric Alternant and Goppa Codes}
\label{sec:symmetries}

The purpose of this section is to recall how quasi-cyclic (QC), quasi-dyadic (QD) and quasi-monoidic (QM) 
alternant/Goppa codes \cite{MB09,BCMN10,BLM11a} and more generally any \emph{symmetric} alternant/Goppa code 
can be constructed from a common principle which stems from
D\"{u}r's work in \cite{Dur87} about the automorphism group of (generalized) Reed-Solomon codes.
This has been applied and developed in 
\cite{Ber00a,Ber00b} to construct large families of symmetric alternant or Goppa codes. It should be emphasized that this way of constructing 
symmetric Goppa codes is more general than the constructions proposed  for \QD{} or \QM{} Goppa in a cryptographic context by \cite{MB09,BCMN10,BLM11a}. 
In particular,  it is required in \cite{MB09,BCMN10,BLM11a} to choose Goppa codes with a separable Goppa polynomial. We will prove in the following that this constraint 
is unnecessary.

In order to recall these results we need a few definitions. 
An {\em automorphism} of a code of length $n$ defined  over $\fq$ is an isometry of the Hamming space $\fq^n$ \textit{i.e.} a linear transform of $\fq^n$ which both preserves the Hamming weight
and leaves the code globally invariant. A well-known fact about such isometries is that they 
consist of permutations and/or non-zero multiplications of the coordinates. 

\medskip

In this paper, we will be interested only in isometries that are permutations. This action is denoted, 
given a permutation $\sigma$ of the symmetric group on $\{0,\dots{},n-1\}$ and a vector $\xv = (x_0,\dots{},x_{n-1})$,
by $\xv^\sigma \eqdef (x_{\sigma(0)},\dots{},x_{\sigma(n-1)})$. For a code $\code{C}$ and a permutation $\sigma$, 
we define:
$$
\code{C}^\sigma \eqdef \left\{\cv^\sigma \mid  \cv \in \code{C}\right\}.
$$
A {\em permutation automorphism} of $\code{C}$ is then any permutation $\sigma$ such that $\cv^\sigma$ is in $\code{C}$ 
whenever $\cv$ belongs to $\code{C}$. {\em Symmetric codes} are then codes with a \emph{non-trivial} automorphism group.

\medskip

We have seen in Proposition~\ref{prop:affinepreserve} that alternant codes may have several identical descriptions thanks to affine transformations.
Actually, symmetric Goppa codes and alternant codes can easily be constructed by looking at the action of the projective semi-linear goup on the 
support of these codes as shown in \cite{Ber00a,Ber00b}. By projective semi-linear group, we mean here transformations of the form:
\begin{eqnarray*}
\fqm \cup \{\infty\} & \rightarrow & \fqm \cup \{\infty\}\\
z & \mapsto&  \frac{az^{q^i}+b}{cz^{q^i}+d}
\end{eqnarray*}
Basically when the support of the alternant code is invariant by the action of such a transformation and under
a certain condition on the multiplier, it turns out that such a transformation induces a permutation automorphism of the 
alternant code. However, this action on the support may transform a coordinate of the support into $\infty$ and a slightly more general 
definition of generalized Reed-Solomon codes and of alternant codes is required to cope with this issue. This is why A. D\"{u}r introduced 
Cauchy codes in \cite{Dur87} which are in essence a further generalization of generalized Reed-Solomon codes. This construction allows to have 
$\infty$ in its support. To avoid such a technicality (and also to simplify some of the statements and propositions obtained here) we will only consider
the subgroup of affine transformations of the projective semi-linear group. It should be noted however that this simplification 
permits to cover all the constructions of symmetric alternant or Goppa codes used in a cryptographic context
 \cite{BCGO09,MB09,BCMN10,BLM11a,Per12}  and in some cases even to generalize them.
 Namely, we will deal with the following cases:
\begin{definition}
Let $\code{C}$ be an alternant or Goppa code defined over a field $\F$ of length $n$, with an automorphism group $\G$. Given a nonnegative integer $\lambda \le n$, we say that $\Cc$ is:
\begin{itemize}
\item Quasi-Cyclic (\QC{}) if $\G$ is of the form $(\Z/\lambda \Z)$,
\item Quasi-Dyadic (\QD{})  if ${\rm char}(\F)=2$ and $\G$ is of the form $(\Z/2 \Z)^\lambda$,
\item Quasi-Monoidic (\QM{})  if $\G$ is of the form $(\Z/p \Z)^\lambda$ with $p={\rm char}(\F)>2$.
\end{itemize}
\end{definition}

Let us now reformulate some corollaries of the results obtained in \cite{Ber00a,Ber00b} in this particular case. 
 The symmetric alternant or Goppa codes that will be obtained here correspond to permutation automorphisms of alternant or Goppa codes  based on the action of affine maps $x \rightarrow a x + b$ on the
support $(x_0,x_1,\dots,x_{n-1})$ of the Goppa code or the alternant code. If this support is globally invariant by this affine map (and $a$ is not equal to $0$), then this induces a permutation $\sigma$ of the code positions $\{0,1,\ldots,n-1\}$ by defining $\sigma(i)$ as the unique integer in $\{0,1,\ldots,n-1\}$ such that $x_{\sigma(i)}=a x_i +b$. In such a case, we say that $\sigma$ is the 
{\em permutation induced by the affine map} $x \rightarrow ax+b$.
Restricting Theorem 1 of \cite{Ber00b} to affine transformations yields immediately

\begin{proposition} \label{pr:description}
Let $a\neq 0$ and $b$ be elements of $\fqm$. Let $\xv \in \fqm^n$ be a support which is globally invariant by the affine map
$x \rightarrow ax+b$. Let  $\sigma$ be the permutation of $S_n$ induced by this affine map. Let $\ell$ be   the order of $\sigma$. Assume that 
 $\yv \in (\F_{q^m})^n$ 
is an $n$-tuple of nonzero elements such that $\exists \, \alpha \in \fqm$ an $\ell$-th root of unity such that
$y_{\sigma(i)} = \alpha y_i$, for all $i \in \{0,1,\ldots,n-1\}$. Then $\sigma$ is a permutation automorphism of the alternant code $\mAlt{t}{\xv}{\yv}$ for any  degree $t>0$.
\end{proposition}
If we want to obtain Goppa codes, we can apply this result and we just have to check that 
the conditions on the support $x_{\sigma(i)}= ax_i+b$ and  multiplier $y_{\sigma(i)} = \alpha y_i$ are compatible with the definition of the 
Goppa code, namely $y_i = \frac{1}{\Gamma(x_i)}$ where $\Gamma(x)$ is the Goppa polynomial. These considerations yield immediately the following corollary of Proposition \ref{pr:description}.
\begin{corollary}
\label{cor:berger}
Let $a\neq 0$ and $b$ be elements of $\fqm$ with $b \neq 0$ when $a=1$.
 Let $\xv \in \Fqm^n$ be a support which is globally invariant by the affine map
$x \rightarrow ax+b$. Let  $\sigma$ be the permutation of $S_n$ induced by this affine map and let  $\ell$ be its order.
Assume that there exists a polynomial $\Gamma(z)$ and an $\ell$-th root of unity $\alpha$ in $\Fqm$ which is such that
\begin{equation} \label{eq:GoppaPolynomial}
\Gamma(az+b) = \alpha \Gamma(z).
\end{equation}
In such a case, $\sigma$ is a permutation automorphism of the 
Goppa code $\Goppa{\xv}{\Gamma}$.
\end{corollary}
This proposition  allows to obtain easily Goppa codes or alternant codes with a non trivial automorphism group that is cyclic.
\begin{remark}
One might wonder whether it is possible to characterize polynomials which satisfy Equation \eqref{eq:GoppaPolynomial}.
In \cite[Theorem 4]{Ber00a} a slightly more general polynomial equation is considered, namely 
$\Gamma(az^{q^s}+b)=\alpha\Gamma(z)^{q^s}$. It is  the particular case of $s=m$ of  Theorem 4 of \cite{Ber00a} 
which is of interest to us here. However, since it deals with the classification of cyclic alternant codes (there is therefore a restriction
on the order compared to the length which trivializes the solutions of this problem in many cases which are of interest to us) and since for further purposes it will be convenient
for us  to remove the assumption
on $\Gamma(z)$ to have no roots in $\{x_0,\dots,x_{n-1}\}$ which is done implicitly in Theorem 4 (and also in Lemma 2 of \cite{Ber00a} 
that is used to prove Theorem 4) we can not use it in our case
directly.
\end{remark} 
The characterization of the solution set to \eqref{eq:GoppaPolynomial} we will use is the following.

\begin{proposition}
\label{pr:PolynomialForm}
Let $\F$ be a field of finite characteristic $p$ and let $a,b,\alpha$ be elements of $\F$, such that (i) $a\neq 0$ and
(ii) $b \neq 0$ when $a=1$.
All the polynomials $\Gamma(z) \in \F[z]$ satisfying
$\Gamma(az+b) = \alpha \Gamma(z)$ have the following form
\begin{itemize}
\item If $a=1$ then necessarily $\alpha=1$, $\ell = p$ and 
$\Gamma(z)$ is any polynomial  in $\F[z]$ of degree a multiple of $p$ which is of the form $\Gamma(z)=P(z^{p}-b^{p-1}z)$.
\item If $a \neq 1$ then there exists a unique integer $d$ in the range $[0,\ldots,\ell-1]$ such that $\alpha=a^d$ and if we denote by $z_0$ the unique fixed point of the affine map $z \rightarrow az+b$, we have that
$\Gamma(z)$ is any polynomial in $\F[z]$ of degree equal to $d$ modulo $\ell$ which is 
 of the form
$(z-z_0)^d P\left((z-z_0)^\ell\right)$. 
\end{itemize}
\end{proposition}
The proof of this proposition can be found in Appendix~\ref{ss:proof_lemma_polynomial}.
By taking polynomials $P$ in this proposition which are such that 
the resulting $\Gamma(z)$ has no zeros in the support $(x_0,\dots,x_{n-1})$ 
we obtain Goppa codes with a cyclic permutation automorphism group. To obtain automorphism groups
which are isomorphic to
$\left(\Z/p\Z\right)^\lambda$, for some $\lambda \ge 1$, we need a slightly more general statement which is the following:

\begin{proposition}
\label{pr:monoidic_alternant}
Let $p \eqdef {\rm char}(\fqm)$. Let 
$\alpha_0,\ldots,\alpha_{\lambda-1} \in \fqm$ be a set of $s$ elements which are $\F_p$-independent over $\fqm$.
Let $G$ be the group of order $p^\lambda$ generated by the $\alpha_i$'s.
Consider a support $\xv \eqdef (x_0,\dots,x_{n-1})$ which is globally invariant by all the affine transformations 
$z \rightarrow z + \alpha_i$ and assume that the multiplier $\yv \eqdef (y_0,y_1,\dots,y_{n-1})$ is constant on the cosets of
$G$ meaning that $y_i = y_j$ iff $x_i - x_j \in G$. Then 
$\Alt{r}{\xv}{\yv}$ is an alternant code with a permutation automorphism group isomorphic
to $\left(\Z/p\Z\right)^\lambda$ for any degree $r$. Let 
$P(z) \eqdef \Pi_{g \in G}(z-g)$, then any polynomial $\Gamma(z)$ of the form
$\Gamma(z) = Q(P(z))$ where $Q$ is a polynomial in $\fqm[z]$ gives a  Goppa
code $\Goppa{\xv}{\Gamma(z)}$ of degree $p^\lambda \deg Q$ with an automorphism group isomorphic to
$\left(\Z/p\Z\right)^\lambda$.
\end{proposition}

\begin{IEEEproof}
All the shifts $z \rightarrow z + \alpha_i$ give rise to a permutation automorphism of the alternant code by 
Proposition \ref{pr:description} and they generate a group of order $p^\lambda$ from the independence assumption on the
$\alpha_i$'s. The statement about Goppa
codes follows by observing that the polynomial $\Gamma(z) = Q\left(\Pi_{g \in G}(z-g)\right)$ is invariant by all the 
shifts $z \rightarrow z + \alpha_i$ and by using Corollary \ref{cor:berger}.
\end{IEEEproof}

\begin{remark}
\begin{enumerate}
\item
A support $(x_0,\dots,x_{n-1})$ satisfying the conditions of Proposition \ref{pr:monoidic_alternant} is easily obtained by taking unions of cosets of 
$G$ and getting a \QD{} or a \QM{} Goppa code is obtained by arranging the support as follows.
We define $\xv=(x_i)_{0 \leq i < n}$ by  choosing elements $x_0,x_{p^\lambda},\ldots,x_{(n_0-1)p^\lambda}$ in different cosets of $\fqm/G$
(where $n=n_0 p^\lambda$).
The remaining $x_i$'s are chosen as follows:
\begin{equation} \label{cons:mono}
x_i = x_{\lfloor i/p^\lambda \rfloor p^\lambda} + \sum_{j=0}^{\lambda-1} i_j \alpha_j.
\end{equation}
It is readily checked that all the \QD{} or \QM{} constructions of 
  Goppa codes of \cite{MB09,BCMN10,BLM11a} are just special cases of this construction. It should be observed that the construction presented
  here is more general. In particular, $\Gamma(z)$  does not need to split over $\fqm$ as in \cite{MB09,BCMN10,BLM11a}.
It may even be irreducible as shown by the example $p=q=2$, $G=\F_2$, $m$ odd and $\gamma(z)=1+z$.
\item By using our proof technique of Proposition \ref{pr:PolynomialForm} it can actually be shown that all polynomials $\Gamma(z)$ invariant by the
shifts $z \rightarrow z + \alpha_i$ are actually polynomials of the form $Q\left(\Pi_g(z-g)\right)$.
\end{enumerate}
\end{remark}
From now on, we will say that the permutation automorphism group of an alternant code or a Goppa code that is obtained by such affine maps
(be it a single affine map or a collection of them) is the {\em permutation group  induced} by such affine maps.
As observed in \cite{Ber00b}, an alternant code or a Goppa code can be invariant by a permutation which is not induced by an affine map
or more generally by an element of the projective semilinear group.  However, there is no general way of constructing this kind of permutation and it 
should also be noted that in the case of GRS or Cauchy codes, the whole permutation group is actually induced by the projective linear group, i.e. the set of 
transformations of the kind $z \rightarrow \frac{az+b}{cz+d}$ (this is actually a consequence of
Theorem 4 of \cite{Dur87}).

%% file: invariant_polynomial.tex
\section{Affine-Invariant Polynomials}
\label{sec:invariant_polynomial}

The key ingredient which  allows to reduce to smaller alternant codes or Goppa codes when these are either quasi-monoidic or 
quasi-cyclic is a fundamental result on the form taken by
polynomials which are invariant by an affine map.
These polynomials will arise as sums of the form:
\begin{equation}
\label{eq:sum_form}
Q(z) \eqdef \sum_{i=0}^{\ell -1} \alpha^iP(\sigma^i(z))
\end{equation}
 where 
$P$ is a polynomial, $\sigma$ an affine map of order $\ell$ and $\alpha$ an $\ell$-th root of unity. Such polynomial sums clearly satisfy  polynomial
Equation \eqref{eq:GoppaPolynomial}, since:
\begin{eqnarray*}
Q(\sigma(z)) & = & \sum_{i=0}^{\ell -1} \alpha^iP(\sigma^{i+1}(z)) 
 =  \frac{1}{\alpha} \sum_{i=0}^{\ell -1} \alpha^{i+1}P(\sigma^{i+1}(z))\\
& = & \frac{1}{\alpha} \sum_{i=0}^{\ell -1} \alpha^{i}P(\sigma^{i}(z))
 =  \frac{1}{\alpha} Q(z).
\end{eqnarray*}

Proposition \ref{pr:PolynomialForm} characterizes all solutions of the polynomial Equation \eqref{eq:GoppaPolynomial}.
Conversely, and this will be crucial in our context, it turns out that all these solutions are of the
form \eqref{eq:sum_form}. To formalize this point, we introduce the following notation
\begin{notation}
Let $I^{\sigma,\alpha}_{\leq t}[z] \subseteq \F_{\leq t}[z]$ be the set  of polynomials of degree $\leq t$ which satisfy \eqref{eq:GoppaPolynomial}, i.e.  which satisfy $P\big(\sigma(z)\big) = \alpha P(z)$. When $\alpha=1$ we will simply write
$I^\sigma_{\leq t}[z]$. Finally, when $t <0$ we adopt the convention that 
$I_{\leq t}[z] = I_{\leq t}^{\sigma,\alpha}[z] = \{0\}$.
\end{notation}

We will first consider the case when $\alpha=1$ and $\sigma(x)=x+b$.
\begin{lemma} \label{lem:sum_is_onto}
Let $\F$ be a field of characteristic $p$.
Let $b$ be a non zero element of $\F$ and denote by $\sigma$ the  shift $\sigma : x \mapsto x +b$. 
Denote by $S$ the mapping defined by:
\begin{eqnarray*}
S :\F[z] & \rightarrow & \F[z]\\
P(z) & \mapsto & \sum_{i=0}^{p-1} P(\sigma^i(z)) 
\end{eqnarray*} 
We have for every nonnegative integer $t$:
\begin{eqnarray}
S\left( \F_{\leq t}[z]\right) & = &I^\sigma_{\leq \left\lfloor \frac{t-p+1}{p} \right\rfloor p}[z]\nonumber \\
& = & \left \{ P(z^p-b^{p-1}z) \mid \deg P \leq \left\lfloor \frac{t-p+1}{p} \right\rfloor \right\}\label{eq:image}
\end{eqnarray}
\end{lemma}

The proof of this lemma can be found in Appendix \ref{ss:proof_lemma_sum_is_onto}.
A similar result holds for affine maps of the form $\sigma(x)=ax+b$ where $a \neq 1$.
\begin{lemma}\label{lem:modified_sum_is_onto} Let $\F$ be a finite field.
Let $a$ be an element of order $\ell \neq 1$ in $\F$, $b$ be an arbitrary element of $\F$, $\sigma$ be the affine map $x \mapsto ax+b$, $d$ be an integer in the range
$[0,\ldots,\ell-1]$ and let $\alpha \eqdef a^d$. We define $S$ by
\begin{eqnarray*}
S :\F[z] & \rightarrow & \F[z]\\
P(z) & \mapsto & \sum_{i=0}^{\ell-1} \alpha^{i} P(\sigma^i(z)) 
\end{eqnarray*} 
If we denote by $z_0$ the unique fixed point of $\sigma$, we have:
\begin{eqnarray}
S\left( \F_{\leq t}[z]  \right) & = & I^{\sigma,\alpha}_{\leq t}[z] \\
& = & \left\{(z-z_0)^d P((z-z_0)^\ell) \mid \deg P \leq \lfloor \frac{t-\ell+d}{\ell}\rfloor \right\},
\end{eqnarray}
\end{lemma}

The proof of this lemma can be found in Subsection \ref{ss:proof_lemma_modified_sum_is_onto} of the
appendix.

%% file: reduction.tex
\section{Reducing to a Smaller Alternant or Goppa Code} 
\label{sec:reduction}

\subsection{Folded codes}
Alternant codes and Goppa codes in particular with a certain non-trivial automorphism group (as considered in Proposition \ref{pr:description})
meet a very peculiar property. Namely it is possible
to derive a new alternant (or a Goppa code) with smaller parameters by simply summing up the coordinates. To define this new code more precisely, we introduce the 
following operator. 
\begin{definition}[Folded code]\label{def:folded}
Let $\code{C}$ be a code and $\G$ be a subgroup of permutations of the set of code positions of $\Cc$.
For each orbit $\G(i) \eqdef \{\sigma(i):\sigma \in \G \}$ we choose 
one representative (for instance the smallest one). Let  $i_0,i_1,\dots,i_{s-1}$  be the set of these representatives.
The {\rm folded code} of $\code{C}$ with respect to $\G$, denoted 
by $\cond{\G}{\code{C}}$, is a code of length $s$ which is given by the set of
words $\overline{\cv}^\G \eqdef\big(\sum_{\sigma \in \G}  c_{\sigma(i_j)} \big)_{0 \leq j \leq s-1}$, where $\cv$ ranges over $\code{C}$.
When $\G$ is generated by a single element $\sigma$, that is $\G = <\sigma>$, we will simply write $\cond{\sigma}{\code{C}}$
instead of $\cond{<\sigma>}{\code{C}}$ and $\overline{\cv}^\sigma$ instead of $\overline{\cv}^{<\sigma>}$.
\end{definition}
This folded code is related to constructions which were considered in the framework of decoding codes
with non-trivial automorphism group \cite{Leg11a,Leg12a}. The approach there was to consider for a code $\code{C}$ 
with non-trivial permutation automorphism $\sigma$ of order $\ell$ (which was supposed to be of order $\ell=2$ in \cite{Leg11a,Leg12a}, but
their approach generalizes easily to other orders) the $\sigma$-subcode $\csigma{\sigma}{\Cc}$ obtained as follows:
$$
\csigma{\sigma}{\Cc} \eqdef \left\{ \cv + \cv^\sigma + \cdots + \cv^{\sigma^{\ell-1}} \mid \cv \in \code{C}  \right\}.
$$
If we denote by $\widetilde{\cv}^\sigma \eqdef \cv + \cv^\sigma + \cdots + \cv^{\sigma^{\ell-1}}$ then it turns out that $\widetilde{\cv}^\sigma$ takes on
a constant value on the orbit $i, \sigma(i),\sigma^2(i),\ldots$ of any code position $i$ that is precisely
the term $\sum_{t=0}^{\ell-1}  c_{\sigma^{t}(i)}$ which appears in the definition of the folded code. Stated differently, 
the words of $\csigma{\sigma}{\Cc}$ are nothing but the words of $\cond{\sigma}{\code{C}}$ where each code coordinate $\bar{c^\sigma_i}$ 
of the latter code is repeated as many times as the size of the orbit of $i$ under $\sigma$. These two codes have therefore the same dimension,
but their lengths are different : the first one has the same length as $\Cc$ whereas the latter has length $s$ (the number of orbits under $\sigma$).

The point of considering such a code for decoding $\Cc$ lies in the fact that 
$\csigma{\sigma}{\Cc}$ is a subcode of $\Cc$ which is typically of much smaller dimension than $\Cc$.
Under mild assumptions, it can be shown that the dimension gets reduced by the order of $\sigma$. More precisely:
\begin{proposition}
\label{pr:dim_folding}
Let $\Cc$ be a code of length $n$ that has a permutation automorphism group  $\G$ of size  $\ell$
and a generator matrix $\Gm$ such that if $g_i$ is a row of
$\Gm$ then $g_i^\sigma$ is also a row of $\Gm$ for any $\sigma \in \G$. Denote by $\{\gv_0,\dots,\gv_{k-1}\}$ the
set of rows of $\Gm$.  Consider the group action of  $\G$ on
the set $\{\gv_0,\dots,\gv_{k-1}\}$ of rows of $\Gm$ where  $\sigma$ acts on $\gv_j$ as $\gv_j \mapsto \gv_j^{\sigma}$
for $\sigma \in \G$.
Assume that  the size of each orbit is equal to $\ell$.
Then, the dimension $\csigma{\G}{\Cc}$ is equal to
$\frac{\dim(\Cc)}{\ell}$. This is also the  dimension of $\cond{\G}{\Cc}$ and the length of this code
is equal to $\frac{n}{\ell}$.
\end{proposition}
\begin{IEEEproof}
This follows at once from the fact that $\csigma{\G}{\Cc}$ is generated by the set of $\widetilde{\gv_i}^{\G} \eqdef \sum_{\sigma \in \G} \gv_i^\sigma$
where the $\gv_i$'s are representatives of each orbit of $\G$ acting on $\{\gv_0,\dots,\gv_{k-1}\}$. These vectors are clearly independent and there
are $\frac{\dim(\Cc)}{\ell}$ such representatives. This implies that the dimension of $\csigma{\G}{\Cc}$ is equal to $\frac{\dim(\Cc)}{\ell}$.
This is also clearly the dimension of $\cond{\G}{\Cc}$ and the length of the latter code is equal to $\frac{n}{\ell}$.
\end{IEEEproof}
\begin{remark}
A generator matrix of this form is precisely what is achieved by all the constructions of monoidic alternant/Goppa/Srivastava
codes proposed in \cite{BCGO09,MB09,BLM11a,BCMN10,Per12}. 
\end{remark}
This can be used to decode a word $\yv$ by decoding instead
$\widetilde{\yv}^\sigma$ in $\csigma{\sigma}{\Cc}$. The point is that this decoding
can be less complex to perform than decoding $\yv$ directly
and that the result of the decoding can be useful to solve the original decoding problem, see
\cite{Leg12a}.

\subsection{Folding alternant codes with respect to a cyclic group}

If we consider the monoidic alternant or Goppa codes constructed in 
\cite{BCGO09,MB09,BLM11a,BCMN10} they have typically
length of the form $n=n_0 \ell$, degree of the form
$r = r_0 \ell$ and dimension of the form
$k = n - rm = \ell(n_0 - r_0 m)$ where $m$ is the 
extension degree of the alternant/Goppa code and $\ell$ is 
the size of the automorphism group of the code.
The automorphism group of these codes satisfies the assumptions of
Proposition \ref{pr:dim_folding} and therefore the folded code
has length $n_0$ and dimension $n_0 - r_0 m$. This 
could suggest that these codes are alternant or Goppa codes
of length $n_0$ and degree $r_0$. In all our experiments we have
noticed that this was indeed the case.
We have proved in \cite{FOPPT13} a slightly weaker result, namely that in the
case of a Goppa code obtained from the constructions of \cite{MB09,BLM11a,BCMN10},
the folded code is included in a Goppa code of length $n_0$ and degree $r_0$.
We will prove a significantly stronger result here, by considering instead the dual of these codes.
It will turn out that the folded dual of those alternant or Goppa codes
will be duals of alternant or Goppa codes and this even if the degree is not of the
form $r_0 \ell$. More precisely, we have:
\begin{theorem}\label{th:compressed_alternant}
Consider an alternant code $\mAlt{t}{\xv}{\yv}$ over $\fq$ of length $n$ with support $\xv=(x_0, x_1,\dots,x_{n-1}) \in \fqm^n$ and multiplier $\yv \in \fqm^n$
with a non trivial permutation automorphism group induced by  the affine map
$x \rightarrow ax+b$ where $a,b \in \fqm$ are such 
that $a \neq 0$ and $b \neq 0$ when $a=1$. 
Denote by $\sigma$ the permutation of $S_n$
induced by this affine map. Let $\ell$ be the order of $\sigma$. By definition of an affine induced automorphism, there exists $ \alpha \in \fqm$ an $\ell$-th root of unity such that $y_{\sigma(i)} = \alpha y_i$ for all $i \in \{0,1,\ldots,n-1\}$. We denote by $d$ the 
integer in $\{0,1,\ldots, \ell-1\}$  verifying $\alpha=a^d$.
Let  
 us denote by $u_0$  the unique fixed point in $\fqm \cup \{\infty\}$ of this affine map. Moreover, we assume that 
$u_0 \not \in \{x_0, x_1,\dots,x_{n-1}\}$. 
In such a case, the action of $\sigma$ on $\{0,1,\dots,n-1\}$ has $\frac{n}{\ell}$ orbits, each of them being of size $\ell$.
Choose a representative $i_0,i_1,\dots,i_{n/\ell-1}$ in each of these orbits.   
There exists $\yv' \in \fqm^{n/\ell}$ and a integer $r$ such that 
$\cond{\sigma}{\left(\mAlt{t}{\xv}{\yv}^\perp\right)}  = \Big(\mAlt{r}{\xv'}{\yv'}\Big)^\perp$ with:
\begin{itemize}
\item when $a = 1$ then $r = \left \lfloor\frac{t-\ell}{\ell} \right \rfloor+1$ and for all $j \in \{0,\ldots,n/\ell-1\}$:
$$
x'_{j}  = 
x_{i_{j}}^\ell-b^{\ell-1} x_{i_{j}} 
~~~~\text{ and }~~~~
y'_{j}  =  y_{i_{j}}  
$$

\item and when $a \neq 1$ then $r =  \left \lfloor\frac{t-\ell+d-1}{\ell} \right \rfloor+1$ and for all 
$j \in \{0,\ldots,n/\ell-1\}$:
$$
x'_{j}  =  (x_{i_{j}}-u_0)^\ell 
~~~~\text{ and }~~~~
y'_{j}  = y_{i_{j}}(x_{i_j}-u_0)^{\ell -d} 
$$
\end{itemize}

\end{theorem}
\begin{IEEEproof}

\paragraph*{\bf The case $a=1$}
remark first that the order $\ell$ of the permutation $\sigma$, which is the shift $x \mapsto x +b$ in this case,
is necessarily the characteristic $p$ of $\fqm$. Since the order of the multiplicative group of $\fqm$, which is $q^m-1$, is coprime with the characteristic of $\fqm$ it follows that $\alpha$ is necessarily equal to $1$ when $a=1$. 
This implies that $\yv$ is constant over each orbit $\{i,\sigma(i),\dots,\sigma^{\ell -1}(i)\}$.
From Lemma \ref{lem:trace}, the dual $\code{C}$ of $\mAlt{t}{\xv}{\yv}$ is:
\begin{equation*}
\label{eq:traceC}
\code{C} = \left\{ (\tr \big(y_i P(x_i)\big)_{0 \leq i < n} \mid P \in \fqm[z], \deg P  \leq t-1 \right\}.
\end{equation*}
The folded code of $\code{C}$ can now be described as:
\begin{equation*}
\cond{\sigma}{\code{C}}= \left\{ \tr \left(y_{i_j} \sum_{s=0}^{\ell -1} P\left(\sigma^s(x_{i_j})\right)\right)_{j=0}^{n/\ell-1} \mid  P \in \fqm[z], \deg P \leq t-1  \right\}
\end{equation*}
where $x_{i_0},x_{i_1},\dots,x_{n/\ell-1}$ are representatives of each of the $n/\ell$ orbits $\{u,\sigma(u),\dots,\sigma^{\ell-1}(u)\}$ (they have all the same size $\ell$).

By using Lemma \ref{lem:sum_is_onto},
we 
obtain:
 \begin{equation}
\cond{\sigma}{\code{C}} = \left\{\tr \left(y_{i_j} R\left(x_{i_j}^p - b^{p-1} x_{i_j} )\right)\right)_{j=0}^{n/\ell-1} \mid R \in \fqm[z], \deg R \leq  \left\lfloor \frac{t-p}{p} \right\rfloor \right\}
\end{equation}
By using Lemma \ref{lem:trace} again, we see that
$
\cond{\sigma}{\code{C}} = \mAlt{r}{\xv'}{\yv'}^\perp
$
with $r = \left\lfloor \frac{t-p}{p}\right\rfloor+1$ and  for any $j \in \{0,1,\ldots,n/\ell-1\}$,
 $x'_{j}  =  x_{i_{j}}^p-b^{p-1} x_{i_{j}}$ and $y'_{j} = y_{i_{j}}$.

\medskip

\paragraph*{\bf The case $a \neq 1$} 
the difference with the previous situation lies in the fact that now the $y_j$'s are not necessarily constant over an orbit. As previously, we consider representatives $x_{i_0},x_{i_1},\dots,x_{n/\ell-1}$ of the $n/\ell$ orbits $\{u,\sigma(u),\ldots,\sigma^{\ell-1}(u)\}$ (they have here again all the same size $\ell$
because the support $\xv$ does not contain the fixed point of $\sigma$).  We obtain that the folded code of $\code{C}$ can now be described as follows.
\begin{equation*}
\cond{\sigma}{\code{C}}= \left\{\tr \left( \sum_{s=0}^{\ell -1} y_{i_j} \alpha^s P\left(\sigma^s(x_{i_j})\right)\right)_{j=0}^{n/\ell-1} \mid  P \in \fqm[z], \deg P \leq t-1  \right\}.
\end{equation*}
By introducing the fixed point $u_0$ of $\sigma$, we obtain:
\begin{eqnarray*}
\cond{\sigma}{\code{C}}&=&\left\{\tr \left( y_{i_j} \sum_{s=0}^{\ell -1}  \alpha^s P\left(u_0 + a^s (x_{i_j}-u_0)\right)\right)_{j=0}^{n/\ell-1}\mid P \in \fqm[z], \deg P \leq t-1  \right\}\\
& = &\left\{\tr \left( y_{i_j} \sum_{s=0}^{\ell -1}  \alpha^s Q\left(a^s (x_{i_j}-u_0)\right)\right)_{j=0}^{n/\ell-1}\mid  Q \in \fqm[z], \deg Q \leq t-1  \right\}.
\end{eqnarray*}
We necessarily have $\alpha^\ell=1$. Since $a$ is a primitive $\ell$-root of unity, there exists an integer 
$d$ in $\{0,\ldots,\ell-1\}$ such that $\alpha=a^d$. This yields:
$$
\cond{\sigma}{\code{C}} = \left\{\tr \left( y_{i_j} \sum_{s=0}^{\ell -1}  a^{ds} Q\left(a^s (x_{i_j}-u_0)\right)\right)_{j=0}^{n/\ell-1}\mid Q \in \fqm[z], \deg Q \leq t-1  \right\}.
$$
By using Lemma \ref{lem:modified_sum_is_onto}, we deduce that:
$$
\cond{\sigma}{\code{C}} = \left\{\tr \left( y_{i_j} (x_{i_j}-u_0)^{\ell - d}  R\left((x_{i_j}-u_0)^\ell\right)\right)_{j=0}^{n/\ell-1}\mid R \in \fqm[z], \deg R \leq \left\lfloor \frac{t-1-\ell + d}{\ell} \right\rfloor  \right\}.
$$
Finally, by Lemma \ref{lem:trace} again we see that
$
\cond{\sigma}{\code{C}} = \mAlt{r}{\xv'}{\yv'}^\perp
$
where $r = \left\lfloor \frac{t-1-\ell + d}{\ell} \right\rfloor+1$, $x'_{j}  =  (x_{i_j}-u_0)^\ell$ and $y'_{j} = y_{i_{j}}(x_{i_j}-u_0)^{\ell - d}$ for any $j \in \{0,1,\ldots,n/\ell-1\}$.
\end{IEEEproof}

\begin{remark}

In essence, we have proved here that folding a GRS code
with a non trivial automorphism group obtained from affine transformations yields again 
a GRS code. Indeed, the dual of an alternant code is the trace of a GRS code. When we choose the extension degree to be equal to $1$ 
we really prove here that folding such a symmetric GRS code  yields again a GRS code. Taking the trace preserves this property :
the folding of a  trace of a symmetric GRS  code is again the trace of a GRS code.
The crucial point which explains why such a property holds is the fact that the ring of polynomial in $\F[x]$ invariant by an affine transformation
$\sigma$  is
a ring of the form $\F[Q(x)]$ for some polynomial $Q$ which is invariant by $\sigma$.
This is what allows to write a sum of the form $\sum_{i=0}^{\ell-1} P(\sigma^i(x))$ as a polynomial
of the form $R(Q(x))$. 

\bigskip

One might wonder whether folding a subfield subcode of a GRS code (i.e. an alternant code) also yields a subfield subcode of
a GRS code. While the proof technique used here obviously allows to prove that a folded subfield subcode of a symmetric
GRS code lies in a subfield subcode of a certain subcode, proving equality of both codes seems to be more delicate here.
This point can be explained as follows. Consider an alternant code $\Alt{r}{\xv}{\yv}$ defined over $\Fq$ and of extension 
degree $m$ where $\xv$ is globally invariant by some $\sigma$ 
and $\yv$ is constant on the orbits on $\sigma$ (we make this assumption to simplify the discussion).
To prove equality that the folded alternant code is still an alternant code we should be 
able to express a polynomial  $Q(z)$ in $\Fqm[z]$  which is invariant by $\sigma$ and which is such that $y_i Q(x_i)$ belongs to 
$\Fq$ for any $i$ as a sum
$Q(x) = \sum_{j=0}^{\ell-1} P(\sigma^{j}(x))$ where all the $y_i P(\sigma^j(x_i))$ belong to
$\Fq$ for any $i$ and $j$ and where $P$ is some polynomial which depends on $Q$. 

\end{remark}

\subsection{Folding alternant codes with respect to non-cyclic groups}

We have treated the case of folding an alternant code with respect to a group generated by a single element. 
The group of automorphism might not be cyclic. This happens in particular in the case of the Goppa codes 
in \cite{MB09,BLM11a,BCMN10}: in such a case the automorphism group 
is isomorphic to $(\Z/p\Z)^\lambda$. Treating the general case of a subgroup of the affine subgroup is beyond the scope of this article, we will just consider the case
of a subgroup which is isomorphic to $(\Z/p\Z)^\lambda$. This follows at once from Theorem \ref{th:compressed_alternant} by noticing that we may fold iteratively 
the code with respect to $\lambda$ generators of the subgroup and end up with an alternant code. We use here the straightforward fact
\begin{lemma}\label{lem:folding_abelian_group}
Consider a code $\Cc$ and a group of permutations $\G$ acting on the positions of $\Cc$ and assume that this permutation group has a 
subgroup $\G_0$ and an element $\sigma$ of $\G$ which does not belong to 
$\G_0$ such that:

\begin{enumerate}
\item the cosets $\sigma^i\G_0$ form a partition of $\G$ for $i \in \{ 0,\dots,\ell-1\}$ where $\ell$ is the
order of $\sigma$;\\
\item $\sigma$ commutes with any element of $\G_0$.
\end{enumerate}

\bigskip

Then $\sigma$ induces a permutation on the set of positions of $\cond{\G_0}{\Cc}$ that we call
$\hat{\sigma}$ which is defined as follows. We view a code position $i$ of $\cond{\G_0}{\Cc}$ as 
an orbit  $\{\tau(u),\tau \in \G_0\}$ for some code position $u$ of $\Cc$ and $\hat{\sigma}(i)$ is given by 
the orbit $\{\tau(\sigma(u)),\tau \in \G_0\}$.
If the order of $\hat{\sigma}$ is equal to the order $\ell$ of $\sigma$ and for an appropriate order on the
choices of the representatives for the orbits under $<\hat{\sigma}>$, $\G_0$ and $\G$, we have
$$
\cond{\hat{\sigma}}{\left(\cond{\G_0}{\Cc}\right)} = \cond{\G}{\Cc}.
$$  
\end{lemma}

\begin{IEEEproof}
First we have to check that the definition of $\hat{\sigma}(u)$ makes sense, i.e. that it does not 
depend on the choice of $u$ in the orbit $\{\tau(u),\tau \in \G_0\}$. This follows from the fact that 
$\sigma$ commutes with any element of $\G_0$. Indeed assume that we have:
$$
\{\tau(u),\tau \in \G_0\} = \{\tau(v),\tau \in \G_0\} 
$$
then we clearly have $u = \tau_0(v)$ for a certain $\tau_0$ in $\G_0$. From that we deduce:
\begin{eqnarray*}
\{\tau(\sigma(u)),\tau \in \G_0\} & = & \{\tau(\sigma(\tau_0(v))),\tau \in \G_0\} \\
& = & \{\tau(\tau_0(\sigma(v))),\tau \in \G_0\}\\
& = & \{\tau(\sigma(v)),\tau \in \G_0\}
\end{eqnarray*}
This shows that $\hat{\sigma}$ is well-defined. We let $i_0,i_1,\dots,i_{s-1}$ be a set of representatives of each orbit 
of the code positions of $\Cc$ under $\G_0$ (we assume that there are $s$ orbits) and we assume
that the set of code positions $0,1,\dots,s-1$ of $\cond{\G_0}{\Cc}$ corresponds to 
$i_0,i_1,\dots,i_{s-1}$ in this order.
Consider now an element $\cv$ in $\Cc$ and let $\cv'$ be the 
folding of $\cv$ with respect to $\G_0$, that is: 
\begin{equation}\label{eq:folding1}
c'_j = \sum_{\tau \in \G_0} c_{\tau(i_j)}
\end{equation}
If we fold $\cv'$ with respect to $\hat{\sigma}$ we obtain an element $\cv''$ defined by:
\begin{equation}\label{eq:folding2}
c''_j = \sum_{l=0}^{\ell-1} c'_{\hat{\sigma}^l(i'_j)}
\end{equation}
where $i'_0,i'_1,\dots,i'_{t-1}$ are the 
representatives of the orbits of the code positions of $\cond{\G_0}{\Cc}$ under $\hat{\sigma}$.
Notice that we have used here the fact that the order of $\hat{\sigma}$ is equal to the order
of $\sigma$.
By observing that the code position $i'_j$ of $\cond{\G_0}{\Cc}$ corresponds to some orbit $\{\tau(u),\tau \in \G_0\}$ and 
putting \eqref{eq:folding1} and \eqref{eq:folding2} together with the characterization of the action of $\hat{\sigma}$, we obtain:
$$
c''_j = \sum_{l=0}^{\ell-1} \sum_{\tau \in \G_0} c_{\tau(\sigma^l(u))} = \sum_{\tau \in \G} c_{\tau(u)}.
$$
This implies that $c''_j$ is equal to some coordinate of $\cond{\G}{\cv}$.

It remains to show that there is a one-to-one and onto mapping from the set of coordinates of $\cv''$ and those of 
$\cond{\G}{\cv}$. In order to do so we are going to prove that there is a one-to-one mapping
between the orbits under $\hat{\sigma}$ and the orbits under $\G$. This is a straightforward consequence of the following observation. Consider an orbit $\Oscr=\{\tau(s),\tau \in \G\}$ under $\G$.
It decomposes as a union of orbits $\Oscr_h$ under $\G_0$: $\Oscr = \cup_{0 \leq h \leq \ell-1} \Oscr_h$ 
where $\Oscr_h \eqdef \{\tau(\sigma^h(s))\}$. These orbits $\Oscr_h$ form a single orbit under $\hat{\sigma}$ and we are done.
\end{IEEEproof}

A straightforward consequence of this is the following
\begin{corollary}
\label{cor:folding_alternant}
Consider a code $\Cc$ which is the dual of an alternant code with an affine-induced permutation group $\G$ 
isomorphic to
$\left(\Z/p\Z\right)^\lambda$ where $p$ is the characteristic of the field over which the alternant code is defined. Then $\cond{\G}{\Cc}$ is the dual of an alternant code.
\end{corollary}

\begin{IEEEproof}
In such a case, there exists $g_1,\dots,g_\lambda$ of order $p$ that generate $\G$.
We proceed by induction and assume that this property holds for $\lambda=h$. When $h=1$, this is just Theorem \ref{th:compressed_alternant}. Consider now a group $\G$ isomorphic to $\left( \Z/p\Z\right)^{h+1}$.
We observe that $\G_0 \eqdef <g_1,\dots,g_{h}>$ and $\sigma=g_{h+1}$
satisfy the assumptions of Lemma \ref{lem:folding_abelian_group}, so we can apply it to this case and obtain that:
$$
\cond{\hat{\sigma}}{\left(\cond{\G_0}{\Cc}\right)} = \cond{\G}{\Cc}.
$$ 
Since by induction
hypothesis $\cond{\G_0}{\Cc}$ is the dual of an alternant code and since $\hat{\sigma}$ is clearly an affine induced permutation
automorphism of $\cond{\G_0}{\Cc}$ we can apply Theorem \ref{th:compressed_alternant} to it and obtain that the result of the
folding of $\cond{\G_0}{\Cc}$ by $\hat{\sigma}$ gives an alternant code again.
\end{IEEEproof}

All the duals of the codes used in the following variants of the McEliece cryptosystem, namely the dyadic Goppa
codes of \cite{MB09,BCMN10}, the monoidic Goppa codes of \cite{BLM11a} or the dyadic Srivastava codes of
\cite{Per12} are instances of alternant codes which have an affine induced permutation group isomorphic to
$\left(\Z/p\Z\right)^\lambda$ and this corollary can be applied to reduce attacks on the key to a much smaller
key recovery problem (namely on the dual of the code obtained by folding). One might also wonder when we fold
certain subfamilies of duals of alternant codes with respect to an affine-induced permutation automorphism group, such as duals of Goppa codes, we stay in the subfamily, i.e. \emph{do we still obtain the dual of a Goppa code?} This turns out to be the case as shown by the next subsection.

\subsection{Folding Goppa codes }
Folding the dual of a Goppa code with an affine-induced automorphism group  yields the dual of an  alternant code 
by using Corollary \ref{cor:folding_alternant}. It turns out that a stronger statement holds: we actually obtain the dual of a Goppa code, both in the
cyclic  case as shown by the following 
theorem and when the group is isomorphic to $\left(\Z/p\Z\right)^\lambda$ as shown later on.
\begin{theorem}
\label{th:folding_Goppa_cyclic}
Consider a Goppa code $\Cc=\mgoppa{\xv}{\Gamma(z)}$ of length $n$ associated to the support $\xv = (x_i)_{0 \leq i < n} \in \fqm^n$ which has a cyclic affine induced automorphism group generated by 
$\sigma(x) \eqdef ax+b$ where $a,b \in \fqm$. We assume that $a \neq 0$, $b \neq 0$ if $a=1$, and that the fixed point $u_0$ of $\sigma$ does not belong to 
$\{x_0,\dots,x_{n-1}\}$.
Let $\ell$ be the order of $\sigma$. 
In such a case:
\begin{enumerate}
\item $\ell$ divides $n$ and let $s \eqdef n/\ell$. There are exactly $s$ orbits for the action of $\sigma$ on the code positions. We denote by  $i_0,i_1,\dots,i_{n/\ell-1}$
a set of representatives for each orbit;
\item $\cond{\sigma}{{\left(\Cc^\perp\right)}}$ is the dual of the Goppa code  $\mgoppa{\xv'}{\gamma(z)}$ with:
\begin{eqnarray*}
x'_{j} &=& \left \{ 
\begin{array}{l}
x_{i_{j}}^\ell-b^{\ell-1} x_{i_{j}}\mbox{ when } a=1,\\
(x_{i_{j}}-u_0)^\ell\mbox{ otherwise,}
\end{array}
\right.\\
\Gamma(z) & = & \left \{ 
\begin{array}{l}
\gamma(z^\ell-b^{\ell-1}z)\mbox{ when } a=1,\\
(z-u_0)^d \gamma\left((z-u_0)^\ell \right) \mbox{ otherwise}
\end{array}
\right.
\end{eqnarray*}
\end{enumerate}
where $d$, in the last case,  is the unique  integer in $\{0,\dots,\ell-1\}$ such that 
$\alpha = a^d$ and $\alpha$ is the element of $\fqm$  which satisfies 
the polynomial identity $\Gamma(a z + b) = \alpha \Gamma(z)$.
\end{theorem}

\begin{IEEEproof}
We will distinguish between $a =1$ and $a \neq 1$. In both cases, notice that we can apply 
 Theorem \ref{th:compressed_alternant}  to $\Cc$ which is an alternant code $\Alt{t}{\xv}{\yv}$ where 
 $t$ is the degree of $\Gamma$ and $y_i = \frac{1}{\Gamma(x_i)}$.
 This is a consequence of the definition of a Goppa code with an affine induced automorphism $\sigma(x) = a x +b$ :
 this is a Goppa code obtained from the construction of Proposition \ref{pr:description} and this is precisely what is needed 
 (together with the fact that the support does not contain the fixed point of $\sigma$) for applying 
 Theorem \ref{th:compressed_alternant} to it. In all cases, folding the dual of $\Cc$ gives the dual of an alternant 
 code of the form $\Alt{t'}{\xv'}{\yv'}$ for some integer $t'$ and some $\xv',\yv'$ in $\fqm^s$. Moreover in both cases,
   there 
 exists an $\ell$-th root of $1$ that we denote by $\alpha$ which is such that the
Goppa polynomial satisfies the identity $\Gamma(a z + b) = \alpha \Gamma(z)$. 

\medskip
{\bf Case } $a=1$ :
 $\ell$ is equal to the characteristic $p$ of the field $\fqm$, $\alpha$ is necessarily equal to $1$,
$\Gamma(z)$ is of degree a multiple of $p$ and is of the form $\Gamma(z) = \gamma(z^p-b^{p-1}z)$.
Notice that $\yv$ satisfies:
\begin{eqnarray*}
y_{\sigma(i)} & = & \frac{1}{\Gamma(ax_i+b)}
 =  \frac{1}{\Gamma(x_i)}
 =  y_i
\end{eqnarray*}
and  using Theorem \ref{th:compressed_alternant} gives that $y'_j = y_{i_j}$ and therefore:
\begin{eqnarray*}
y'_{j} & = & y_{i_j} 
 =  \frac{1}{\Gamma(x_{i_j})}
 =  \frac{1}{\gamma(x_{i_j}^p-b^{p-1}x_{i_j})} 
 =  \frac{1}{\gamma(x'_j)}
\end{eqnarray*}
This implies that $\Alt{t'}{\xv'}{\yv'}$ is nothing but the Goppa code $\mgoppa{\xv'}{\gamma(z)}$.

\medskip
{\bf Case } $a \neq 1$ :
there exists a unique integer $d$ in the range $[0,\dots,\ell-1]$ such that 
$\alpha = a^d$ and $\Gamma(z)$ is of the form $\Gamma(z) = (z-u_0)^d \gamma\left((z-u_0)^\ell \right)$.
Notice that in such a case:
\begin{eqnarray*}
\frac{1}{y_{\sigma(i)}} & = & \Gamma(ax_i+b)
 =  (ax_i+b-u_0)^d \;\gamma\left((ax_i-u_0)^\ell \right)\\
& = & (ax_i+b-au_0-b)^d \; \gamma\left((ax_i+b-au_0-b)^\ell \right)\\
& = & (a(x_i-u_0))^d \; \gamma\left(a^\ell (x_i-u_0)^\ell \right)\\
& = & a^{d} (x_i-u_0)^d \; \gamma\left((x_i-u_0)^\ell \right)\\
& = & a^{d}\Gamma(x_i)
 = a^{d} \frac{1}{y_i}
\end{eqnarray*}
We use Theorem \ref{th:compressed_alternant} and obtain:
\begin{eqnarray*}
y'_{j} & = & y_{i_j} (x_{i_j}-u_0)^{d} 
 =  \frac{(x_{i_j}-u_0)^{d}}{\Gamma(x_{i_j})}
 =  \frac{(x_{i_j}-u_0)^{d}}{(x_{i_j}-u_0)^d \gamma\left((x_{i_j} -u_0)^\ell \right)} 
 =  \frac{1}{\gamma(x'_j)}
\end{eqnarray*}
This implies again that $\Alt{t'}{\xv'}{\yv'}$ is nothing but the Goppa code $\mgoppa{\xv'}{\gamma(z)}$.
\end{IEEEproof}

When the group is isomorphic to $(\Z/p\Z)^\lambda$ we have the following statement
\begin{theorem}\label{th:compressed_goppa}
Consider a Goppa code $\Cc = \mgoppa{\xv}{\Gamma}$ with an affine induced automorphism group $\G$  isomorphic to 
$(\Z/p\Z)^\lambda$ where $p$ is the characteristic of the field over which the Goppa code is defined, then the folding $\cond{\G}{{\left(\Cc^\perp\right)}}$ is the dual of a Goppa code  $\mgoppa{\xv'}{\gamma(z)}$
where the degree $\deg(\gamma)$ of $\gamma$ is equal to 
$\frac{\deg(\Gamma)}{p^\lambda}$.
\end{theorem}
\begin{IEEEproof}
We proceed similarly to the proof of Corollary \ref{cor:folding_alternant}.
First we notice that there exists $g_1,\dots,g_\lambda$ of order $p$ that generate $\G$.
We proceed by induction and assume that this property holds for $\lambda=h$. When $h=1$, this is just Theorem \ref{th:folding_Goppa_cyclic} (since 
$g_1$ is necessarily induced by an affine transformation of the form $x \mapsto x + \beta$ which has no fixed point in the extension field in which the coordinates of the multiplier live).
 Consider now a group $\G$ isomorphic to $\left( \Z/p\Z\right)^{h+1}$.
We observe that $\G_0 \eqdef <g_1,\dots,g_{h}>$ and $\sigma=g_{h+1}$
satisfy the assumptions of Lemma \ref{lem:folding_abelian_group}, so we can apply it to this case and obtain that:
$$
\cond{\hat{\sigma}}{\left(\cond{\G_0}{\Cc}\right)} = \cond{\G}{\Cc}. 
$$
Since by induction
hypothesis $\cond{\G_0}{\Cc}$ is the dual of a Goppa code of degree $\frac{\deg(\Gamma)}{p^h}$ and since $\hat{\sigma}$ is clearly an affine induced permutation
automorphism of $\cond{\G_0}{\Cc}$ we can apply Theorem \ref{th:folding_Goppa_cyclic} to it and obtain that the result of the
folding of $\cond{\G_0}{\Cc}$ by $\hat{\sigma}$ gives the dual of a  Goppa code of degree $\frac{\deg(\Gamma)}{p^{h+1}}$.

\end{IEEEproof}

%% file: implications.tex
\section{Conclusion -- Cryptographic Implications}

The results presented on this paper have some significant consequences on 
 a recent research trend which consists in  devising McEliece schemes with reduced  public key size. This is achieved by relying on \QD{}/\QM{} Goppa codes
or \QC{} alternant codes \cite{BCGO09,MB09,BCMN10,BLM11a}. Some of them were attacked by the algebraic attack introduced in \cite{FOPT10,UL09} where it was proved that the 
\QD{} or the \QC{} structure allowed to set up an algebraic system which could be solved by 
Gr\"obner bases techniques 
thanks to the reduction of unknowns obtained in this case compared to an unstructured
McEliece scheme. Our result actually explains where this reduction in the number of unknowns comes from: there is in fact a smaller \emph{hidden} Goppa (or alternant) code behind
the public generator or parity-check matrix of the scheme.  Moreover it is shown in  \cite{FOPPT13}  that a key recovery attack on the reduced cryptosystem can be used
to  recover the secret key of the original cryptosystem. This implies that a key-recovery on \QD{} and \QM{} schemes is not harder than a key-recovery on a reduced McEliece scheme where all parameters have been scaled down by a factor of $p$, which is the compression factor allowed by the \QC{}, \QD{} or \QM{} structure. For instance,  we can reduce the key-recovery of a \QD{} Goppa code of length $8192$ and dimension $4096$ (parameters suggested in \cite{MB09}) to the key-recovery on a  \QD{} Goppa code of length $64$ and dimension $32$. In other words, the very reason which allowed to design compact variants of McEliece can be used to attack such schemes much more efficiently.

\medskip

Our result does not rule out the possibility of devising alternant or Goppa codes with a non trivial automorphism group for which folding does not produce
an alternant or a Goppa code: it only applies to such codes with an affine induced automorphism group. Symmetric codes of this kind could be obtained
from the action of the semi-linear projective group on the support instead of the affine group (see Section \ref{sec:symmetries}). It is an open question to understand if folding 
such symmetric codes yields again Goppa or alternant codes, but obviously even treating the case of the linear projective group (obtained from the transformations of the
kind $z \rightarrow \frac{az+b}{cz+d}$) needs much more general tools than those that have been considered here and is beyond the scope of this paper.
It should also be added that this result does not mean that all compact key McEliece cryptosystems based on alternant or Goppa codes with  an affine induced 
automorphism group are weak. It just means that the key security is not better than the key security of a reduced scheme obtained from the folding process. Since 
key recovery attacks are generally more expensive that message recovery attacks it might be possible to choose secure parameters for which we still obtain a good reduction of the
key size where key recovery attacks on the folded key are of the same complexity as  message recovery attacks on the original scheme.
 However this thread of research requires great care since there has been some recent progress on  key recovery attacks, see \cite{FOPPT13,COT14} for instance.

%% file: proofs.tex

\subsection{Proof of Proposition \ref{pr:PolynomialForm}} \label{ss:proof_lemma_polynomial}

We will first characterize the solutions to Equation \eqref{eq:GoppaPolynomial} in the case where
$\alpha=1$. In some sense, this characterization
 generalizes a classical result about even polynomials, i.e. polynomials $P(z)$ which 
 satisfy $P(z)=P(-z)$. It is namely well known that a polynomial is even if and only if 
 there exists a polynomial $Q$ such that $P(z)=Q(z^2)$. Lemma \ref{lem:polynomial}, which uses the notation 
$I^\sigma_{\leq t}[z]$ that is defined in Section \ref{sec:invariant_polynomial}, will generalize this result to
 any polynomial invariant under a finite order affine map.
 \begin{lemma}
 \label{lem:polynomial}
 Let $\sigma(z)=az+b$ be an affine map of finite order $\ell$ (with $\ell >1$) defined over a field $\F$.
 We have
 \begin{itemize}
 \item  if $a=1$ then $\F$ is of characteristic $\ell$ and $I^\sigma_{\leq t}[z] = \left\{ Q(z^\ell-b^{\ell-1}z) \mid \deg Q \leq t/\ell \right\}$.
 \item if $a\neq 1$ then 
  $I^\sigma_{\leq t}[z] = \left\{ Q((z-z_0)^\ell) \mid \deg Q 
 \leq t/\ell \right\}$ with $z_0$ 
 being the unique fixed point of $\sigma$.
 \end{itemize}
 \end{lemma}

In other words, the ring of polynomials  invariant by an affine map is generated by a single element and the lemma provides
 this generator explicitly. This result follows from classical results in invariant theory and we derive it from scratch here to keep the paper self-contained.  Also, we treat the case where the order $\ell$ of the group generated by $\sigma$ is divisible by  the characteristic of $\F$. This is precisely what happens when $a=1$,
 and that is
 commonly avoided in invariant theory (see for instance \cite[Appendix,\S4,Prop.1]{Shaf94}).
\begin{IEEEproof}[Proof of Lemma \ref{lem:polynomial}]
Let us first prove that the right hand side terms which appear in the expressions for $I_{\leq t}^\sigma[z]$ are indeed 
included in  $I_{\leq t}^\sigma[z]$. If $a=1$, consider a polynomial $P$ of degree $\leq t$ of the form $P(z)=Q(z^\ell-b^{\ell-1}z)$
for some polynomial $Q$.
 We have:
\begin{eqnarray*}
P(z+b) & = & Q\left((z+b)^\ell-b^{\ell-1}(z+b)\right)\\
& = & Q\left(z^\ell + b^\ell - b^{\ell-1} z -b^\ell\right)\\
& = & Q\left(z^\ell-b^{\ell-1} z\right)\\
& = & P(z).
\end{eqnarray*}
We just used the fact that $\ell$ is the characteristic of $\F$ and therefore
$(z+b)^\ell=z^\ell+b^\ell$. 

In the case $a\neq 1$, if we consider a polynomial $P$ of degree $\leq t$ of the form $P(z)=Q\big((z-z_0)^{\ell}\big)$ for some 
polynomial $Q$ of degree $ \deg P/\ell$ we obtain:
\begin{eqnarray*}
P(az+b) & = & Q\left((az+b-z_0)^\ell\right)\\
& = & Q\left((az+b-az_0-b)^\ell\right)\\
& = & Q\left(a^\ell(z-z_0)^\ell\right)\\
& = & Q\big((z-z_0)^\ell\big)\\
& = &P(z).
\end{eqnarray*}
We used the fact that $\ell$ is also the order of $a$.

Let us prove now the reverse inclusion. Let $P$ be a polynomial which is invariant by $\sigma$. 
Consider now a non constant polynomial $R$ of smallest degree which is invariant 
by $\sigma$. Such a polynomial necessarily exists since the set of polynomials which are non constant and which are invariant 
by $\sigma$ is non empty 
(since $z^\ell-b^{\ell-1} z$ in the case $a=1$ and
$(z-z_0)^\ell$ in the case $a \neq 1$, belong to it). 
Perform the division of $P$ by $R$. We can write
\begin{equation}
\label{eq:un}
P(z) = R(z) P_1(z) + P_2(z)
\end{equation}
with $\deg P_2 < \deg R$.
Observe now that 
\begin{equation}
\label{eq:deux}
P(az+b)  = R(az+b)P_1(az+b)+P_2(az+b).
\end{equation}
Since $P(az+b)=P(z)$ and $R(az+b)=R(z)$  we deduce by subtracting the second equation to the first one, that we have
$$
R(z)\left( P_1(az+b)-P_1(z)\right)= P_2(z) - P_2(az+b)
$$
Since the degree of $S(z) \eqdef P_2(z) - P_2(az+b)$ is less than the degree of $R$, this can only happen if $P_1$ is invariant under $\sigma$ 
and therefore also $P_2$. Since $R$ is a non constant polynomial of smallest degree which is invariant under $\sigma$ and since $\deg P_2 < \deg R$, this implies that $P_2$ is constant. 
By carrying on this process (i.e. dividing $P_1$ by $R$) we eventually 
obtain that $P$ is a polynomial in $R$. We finish the proof by proving that $R$ can be chosen to be
$R(z)=z^\ell-b^{\ell-1} z$ in the case $a=1$ and $R(z)=(z-z_0)^\ell$ otherwise.

\medskip

Let us first prove this for $a=1$. We can add any constant to $R$, it will still be invariant under $\sigma$. We may therefore assume that $R(0)=0$. We can also assume that $R$ is monic. 
Let us observe now that $0=R(0)=R(b)=R(2b)=\ldots=R\big((\ell-1)b\big)$ by the invariance of $R$ under $z \mapsto z+b$. This implies that $R$ is a multiple
of $z(z-b)\cdots(z-b (\ell-1))$. $R$ is therefore of degree greater than or equal to $\ell$. The polynomial $z^\ell-b^{\ell-1} z$ is of degree $\ell$, is invariant under $\sigma$
and is a multiple of $z(z-b)\ldots(z-b (\ell-1))$. Therefore $R(z)=z^\ell-b^{\ell-1} z$. 

\medskip

Consider now the case $a \neq 1$. Without loss of generality (by adding a suitable constant as in the case $a=0$) we may assume that 
$R(c)=0$, where $c$ is some element of $\F$ such that the orbit of $c$ under $\sigma$ is of size $\ell$. By the invariance of $R$ under $\sigma$ this implies that 
$0=R(c)=R(\sigma(c))= \dots = R\big(\sigma^{\ell-1}(c)\big)$. This implies that $R(z)$ is divisible by $(z-c)(z-\sigma(c))\cdots\big(z-\sigma^{\ell-1}(c)\big)$. Therefore 
$R$ is of degree $\ell$ at least. Since $(z-z_0)^\ell$ is of degree $\ell$ and is invariant by $\sigma$ we can choose $R(z)=(z-z_0)^\ell$.
\end{IEEEproof}

This proves Proposition \ref{pr:PolynomialForm} when $\alpha=1$. Let us prove now this proposition in general.

\begin{IEEEproof}[Proof of Proposition \ref{pr:PolynomialForm}]
Denote by $\sigma$ the affine map $z \mapsto az+b$.
First of all, let us notice that if there exists some polynomial $P(z)$ satisfying the equation $P(\sigma(z)=\alpha P(z)$ for some $\alpha$, then necessarily such an $\alpha$ satisfies $\alpha^\ell=1$. This follows at once from the fact that we have
$P(z)=P\left(\sigma^\ell(z)\right)=\alpha^d P(z)$. This also implies that the order of $\alpha$ divides $\ell$. 
There are now two cases to consider.

{\bf Case $a=1$:}
 then the order $\ell$ of  $\sigma$ is necessarily equal to the characteristic of $\F$ and there is no element, apart from 
$1$, whose order divides $\ell$. In this case, Lemma \ref{lem:polynomial} implies Proposition \ref{pr:PolynomialForm}. 

{\bf Case $a \neq 1$:}
in such a case the order of $a$ is equal to $\ell$ and $a$ is a primitive $\ell$-th root of unity. Since $\alpha$ is an 
$\ell$-th root of unity, there exists in this case
 an integer $d$ in the range $[0,\dots,\ell-1]$ such that $\alpha = a^d$. Consider now a polynomial which is 
such that 
\begin{equation}
\label{eq:base}
P\left(\sigma(z)\right)=\alpha P(z).
\end{equation}
If $\alpha=1$, then we can use directly Lemma \ref{lem:polynomial} and we are done.
Otherwise, 
observe that from the fact that $\sigma(z_0)=z_0$ we deduce that 
$$
P(z_0) = P\left(\sigma(z_0)\right) = \alpha P(z_0).
$$
This implies that $P(z_0)=0$.
Define now a polynomial $P_1$ by $P(z) = (z-z_0)P_1(z)$. Observe now that on the one hand
$$
P(az+b) = (az+b-z_0) P_1(az+b)=a(z-z_0)P_1(az+b)
$$
and that on the other hand
$$
P(az+b) = \alpha P(z) = a^d (z-z_0) P_1(z).
$$
Putting both equations together, we obtain
$$
P_1(az+b)=a^{d-1} P_1(z)
$$
If $d \neq 1$ we can carry on this process on $P_1$, deduce from the previous equation that $P_1(z_0)=0$ and deduce by induction on $d$ that 
$P(z)$ has a zero of order at least $d$ at $z_0$ and that the polynomial $P_d(z)$ defined by 
$P_d(z) = \frac{P(z)}{(z-z_0)^d}$ satisfies the equation
$$
P_d(az+b) = P_d(z).
$$ 
We apply Lemma \ref{lem:polynomial} to $P_d$ and derive from it that $P$ should be of the form
$$
P(z) = (z-z_0)^d Q\left( (z-z_0)^\ell \right),
$$
where $Q$ is any  polynomial of degree $\frac{\deg P - d}{\ell}$.
Conversely, any polynomial $P$ of this form is readily seen to verify \eqref{eq:base}.

\end{IEEEproof}
\subsection{Proof of Lemma \ref{lem:sum_is_onto}}
\label{ss:proof_lemma_sum_is_onto}

For this result, we will need the following lemma.
\begin{lemma}\label{lem:marrant}
$1^k + 2^k + \dots + (p-1)^k \equiv 0 \pmod{p}$ for every integer $k$ which is not a multiple of $p-1$
whereas  $1^{k} + 2^{k}+\dots+(p-1)^{k} \equiv -1 \pmod{p}$
otherwise.
\end{lemma}
\begin{IEEEproof}
Recall that  the multiplicative group $\F_p^\times$ is generated by a single element $\alpha$ which is of order 
$p-1$. The mapping 
\begin{eqnarray*}
\phi_k : \F_p^\times & \rightarrow & \F_p^\times\\
x & \mapsto & x^k
\end{eqnarray*}
maps therefore $\F_p^\times$ to a subgroup of $\F_p^\times$ different from
the trivial subgroup consisting only of $1$ if and only if $k$ is not 
a multiple of $p-1$. In other words, if $k$ is a multiple of $p-1$, we have $s^k \equiv 1 \pmod p$ for 
any $s \in \{1,\ldots,p-1\}$. This implies that $1^{k} + 2^{k}+\dots+(p-1)^{k} \equiv p-1 \equiv -1 \pmod{p}$.
Assume now that $k$ is not a multiple of $p-1$. Thus $\phi_k(\F_p^\times)$ is a subgroup of $\F_p^\times$ 
of size a divisor $\ell>1$ of $p-1$. Since $\F_p^\times$ is generated by $\alpha$, $\phi_k(\F_p^\times)$ is generated
by $\beta \eqdef \alpha^k$ and we have
\begin{eqnarray*}
1^k + 2^k + \dots + (p-1)^k & \equiv & \frac{p-1}{\ell}  \left( 1 + \beta + \dots + \beta^{\ell-1} \right) \pmod{p}\\
& \equiv &  \frac{(p-1)(\beta^{\ell}-1)}{\ell(\beta-1)}\pmod{p}\\
& \equiv & 0 \pmod{p}
\end{eqnarray*}
\end{IEEEproof}

Let us prove now Lemma \ref{lem:sum_is_onto}.
\begin{IEEEproof}
Let us first compute $S(z^{t})$, where $t$ is some nonnegative integer. 
\begin{eqnarray}
S(z^{t}) & = &\sum_{s=0}^{p-1} (z+sb)^{t} =   z^{t} +  \sum_{s=1}^{p-1} \sum_{i=0}^{t} \binom{t}{i}  z^{t-i}(sb)^{i} 
 =  \sum_{s=1}^{p-1} \sum_{i=1}^{t}  \binom{t}{i}  z^{t-i}(sb)^{i} \nonumber \\
& = &  \sum_{i=1}^{t} b^i  \binom{t}{i}  \left( \sum_{s=1}^{p-1} s^{i}\right) z^{t-i} =  \sum_{i=p-1}^t b^i \binom{t}{i}  \left( \sum_{s=1}^{p-1} s^{i}\right) z^{t-i} \label{eq:calculation}
\end{eqnarray}
where the last equation follows by using Lemma \ref{lem:marrant} which allows us to write 
$\sum_{s=1}^{p-1} s^{i} =0$ when $i$ is in the range $[1..p-2]$ and when the sum is performed over a field of characteristic $p$.
This implies immediately that $S(\F_{\leq t}) \subseteq \F_{\leq t-p+1 }[z]$.
Since $S(Q(z))$ is obviously invariant by $\sigma$  for any polynomial $Q(z) \in \F[z]$, we know from Lemma \ref{lem:polynomial} that it is of the form 
$S(Q(z)) = R(z^p - b^{p-1}z)$ for some polynomial $R$ in $\F[z]$. Its  degree is therefore a multiple of $p$.
This implies that we actually obtain the refined inclusion
\begin{equation}\label{eq:inclusion1}
S(\F_{\leq t}) \subseteq I_{\leq \left\lfloor \frac{t-p+1}{p} \right\rfloor p}[x].
\end{equation}
Equality is proven by dimension considerations. It follows from Lemma \ref{lem:polynomial} that 
$I_{\leq t}[z]$  is a vector space which is of dimension $\lfloor t/p\rfloor +1$. The calculation \eqref{eq:calculation} performed above
also shows that $S(z^{(k+1)p-1})$ is a polynomial of degree $kp$ (since the coefficient
of $z^{kp}$ which is equal to $b^{p-1}\binom{(k+1)p-1}{p-1}   \sum_{s=1}^{p-1}s^{p-1}$  by \eqref{eq:calculation} 
can be shown to be different from $0$ by using the fact proven in Lemma \ref{lem:marrant} which says that $1^{p-1} + 2^{p-1}+\dots+(p-1)^{p-1} \equiv -1 \pmod{p}$).
This can be used to obtain that
$$
\dim S(\F_{\leq t}) \geq  \left\lfloor \frac{t-p+1}{p} \right\rfloor + 1 = \dim I_{\leq \left\lfloor \frac{t-p+1}{p} \right\rfloor p}[z].
$$
This together with \eqref{eq:inclusion1} implies that
$$
S(\F_{\leq t}) = I_{\leq \left\lfloor \frac{t-p+1}{p} \right\rfloor p}[x],
$$
which concludes the proof.
\end{IEEEproof}

\subsection{Proof of Lemma \ref{lem:modified_sum_is_onto}}
\label{ss:proof_lemma_modified_sum_is_onto}
\begin{IEEEproof}
Let us calculate
\begin{eqnarray*}
S(z^t) & = & \sum_{i=0}^{\ell-1} a^{di} \left(a^i(z-u_0)\right)^t,\\
& = & (z-u_0)^t \sum_{i=0}^{\ell-1} a^{(d+t)i}. 
\end{eqnarray*}
This sum is equal to $0$ as long as $d+t \not\equiv 0 \pmod{\ell}$ and 
is equal to $\left( \ell \mod p \right)(z-u_0)^t $ when $d+t \equiv 0 \pmod{\ell}$.
The polynomial $S\big(P(z)\big)$ is therefore a polynomial of degree $\ell - d+\left\lfloor \frac{\deg P-\ell + d}{\ell} \right\rfloor \ell$ of the form
\begin{equation}
\label{eq:modified_sum}
S(P(z)) = (z-u_0)^{l-d} \sum_{i=0}^{\left\lfloor \frac{\deg P-\ell + d}{\ell} \right\rfloor} a_i (z-u_0)^{i \ell}
\end{equation}
when $\deg P \geq \ell -d$ and is equal to zero otherwise.
We conclude the proof by noting that the term $\sum_{i=0}^{\left\lfloor \frac{\deg P-\ell + d}{\ell} \right\rfloor} a_i (z-u_0)^{i \ell}$ is a polynomial which is invariant 
by $\sigma$ by Lemma \ref{lem:polynomial}.
\end{IEEEproof}